\documentstyle[eqsecnum,aps,multicol,epsf,float]{revtex}
\begin{document}
\draft
\title{
Noncollinear cluster magnetism in the framework of the Hubbard model
      }
\author{Miguel A.~Ojeda and J. Dorantes-D\'avila}
\address{
Instituto de F\'{\i}sica, Universidad Aut\'onoma de San Luis Potos\'{\i},\\
Alvaro Obreg\'on 64, 78000 San Luis Potos\'{\i}, Mexico
        }

\author{G. M. Pastor}
\address{
Laboratoire de Physique Quantique, Unit\'e Mixte de Recherche 5626 du CNRS,\\ 
Universit\'e Paul Sabatier, 118 route de Narbonne, F-31062 Toulouse, France
        }
\date{\today}
\maketitle
\begin{abstract}

Noncollinear magnetic states in clusters are studied by using the 
single-band Hubbard Hamiltonian. The unrestricted Hartree-Fock (UHF) 
approximation is considered without imposing any symmetry constraints 
neither to the size or orientation of the local magnetic moments 
$\langle\vec S_l\rangle$ nor to the local charge densities 
$\langle n_l\rangle$. A variety of qualitatively different selfconsistent 
solutions is obtained as a function of cluster size, structure, number of 
valence electrons $\nu$ and Coulomb interaction strength $U/t$. This 
includes inhomogeneous density distributions,  paramagnetic solutions, 
magnetic solutions with collinear moments and noncollinear 
spin arrangements that show complex antiferromagnetic and ferromagnetic-like 
orders. The environment dependence of the magnetic properties is analyzed 
giving emphasis to the effects of antiferromagnetic frustrations in compact 
structures close to half-band filling. 
Electron correlation effects are quantified by comparing UHF and exact
results for the local magnetic moments, total spin,
spin-correlation functions and structural stability of $13$-atom clusters. 
Goals and limitations of the present noncollinear approach are discussed.

\end{abstract}
\maketitle
\begin{multicols}{2}
\narrowtext
\section{Introduction}
\label{sec:intro}

Clusters may show specific phenomena that do not have an equivalent in 
the thermodynamic limit and that are therefore of interest in
both basic and applied science. Moreover, the evolution of their 
physical properties with increasing cluster size provides new
perspectives for understanding the microscopic origin of   
condensed-matter properties \cite{gen_clus}. In this context, 
the study of magnetism in metals posses a particularly interesting 
and challenging problem since the  properties of magnetic metals 
often change dramatically  as the  electrons of an isolated atom become 
part of a cluster of several atoms and delocalize. It is therefore of 
fundamental importance to understand how itinerant magnetism,
as found for example in $3d$ transition-metal (TM) solids, develops 
starting {}from localized atomic magnetic moments. The strong dependence 
of the magnetic behavior as a function of system size, structure and 
composition opens in addition the possibility of using  clusters to 
tailor new magnetic materials for specific technological purposes. 
Consequently, the relation between local atomic environment, cluster 
structure and magnetism has also implications of practical interest.

Most theoretical studies of itinerant cluster magnetism have been 
performed using Hartree-Fock (HF) and local spin density (LSD) 
methods \cite{teo-col}. Exact many-body calculations  are 
presently limited to  simple models, such as the Hubbard model, and 
to systems containing a small number of sites \cite{exact2,prlgus,prbflo}.
Within mean-field approximations (HF or LSD) the self-consistent 
spin polarizations are usually  restricted to be collinear, i.e., the 
direction of all local magnetic moments is assumed to be the same.
However, the  well known sensitivity of itinerant magnetism to the 
local environment of the atoms \cite{teo-col} suggests that other 
instabilities towards spiral-like spin-density-waves (SDW's) \cite{penn} 
or even more complex magnetic structures should also be 
possible in general. For example, the tendency to antiferromagnetic 
(AF) order close to half-band filling and the presence of triangular loops  
in compact structures generates magnetic frustrations that may easily 
yield complex arrangements of the local magnetic moments.
Moreover, the reduction of local coordination numbers at the surface 
of clusters removes constraints between the local moments
and could favor the development of low-symmetry spin 
polarizations. In solids,  noncollinear magnetic structures have been
identified experimentally and theoretically already for a long time 
\cite{penn,exp_nocolin,heine,gen_nocolin}. In contrast, very little is 
known at present in the case of finite systems \cite{c60,maol,fe-nc}. 
Fully unrestricted {\em ab initio} calculations of noncollinear spin
arrangements have been performed  only recently for very small 
Fe$_N$ clusters \cite{fe-nc}. The investigation of magnetic phenomena 
of this kind requires a symmetry unrestricted approach in which no 
{\em a priori} assumptions are made concerning the relative orientation 
of the local magnetic moments, thereby enlarging the number of degrees 
of freedom of the problem. 

The main purpose of this paper is to investigate the characteristics 
of noncollinear magnetic states in finite clusters. 
We consider the single-band Hubbard model and determine
the ground-state magnetic properties  in a fully 
unrestricted Hartree-Fock approximation. The theoretical approach,
outlined in Sec.~\ref{sec:model}, is applied to clusters having $N\le 43$
atoms. The results  presented in Sec.~\ref{sec:nocolin} analyze
noncollinear magnetic behaviors for a few representative compact 
structures. Several examples are given that illustrate the large variety 
of 3-dimensional magnetic arrangements obtained in the self-consistent 
calculations as a function of the Coulomb repulsion strength $U/t$, 
band filling $\nu/N$ and cluster structure. In Sec.~\ref{sec:corr}
we compare the UHF results with  exact diagonalization calculations 
for $N\le 13$ atoms (Lanczos method)\cite{prbflo}. The role of 
quantum fluctuations beyond mean field and the consequences of the often 
artificial breaking of spin-symmetry implied by the formation of the  
noncollinear local moments are discussed. Taking into account that 
approximations such as 
UHF are unavoidable for larger clusters and for more realistic
model Hamiltonians, it is  of considerable interest to test the validity 
of these methods in order to improve the  interpretation of the 
approximate results and to obtain a more  accurate description of 
magnetic  phenomena in clusters. 

%
\section{Theoretical method}
\label{sec:model}

The single-band Hubbard Hamiltonian~\cite{hubbard} is given by
\begin{equation}
 H = -t\sum_{\langle l,m \rangle, \sigma} 
\hat c_{l\sigma} ^\dagger \hat c_{m\sigma}
+U\sum_{l} \hat n_{l\uparrow}\hat n_{l\downarrow}\;,
\label{eq_hamil}
\end{equation}
where $\hat c_{l\sigma}^\dagger$ ($\hat c_{l\sigma}$) is the creation 
(annihilation) operator of an electron at site $l$ with spin $\sigma$, 
and $\hat n_{l\sigma}=\hat c_{l\sigma}^\dagger \hat c_{l\sigma}$ is
the corresponding number operator 
($\hat n_{l}=\sum_\sigma \hat n_{l\sigma}$).
The first sum runs over all pairs of nearest neighbors (NN) and the 
second over all sites. The model is characterized by the dimensionless 
parameter $U/t$, that measures the relative importance between kinetic 
and Coulomb energies, by the cluster structure, that defines the kinetic 
energy operator, and by the number of valence electrons $\nu$.
The variations of $U/t$ can be associated to a uniform relaxation of 
the interatomic distances (e.g., $t \propto R_{ij}^{-5}$ for TM's) or to
changes in the spatial extension of the atomic-like wave function, as 
in different elements within the same group. Different $\nu$'s correspond 
to different band-fillings $\nu/N$ that may be associated qualitatively 
to the variations of $\nu/N$ across a TM $d$ series. In spite of its 
simplicity, this Hamiltonian has played, together with 
related models, a major role in guiding our understanding of the 
many-body properties of metals and of low-dimensional magnetism. It is 
the purpose of this work to use it to investigate the properties 
of noncollinear itinerant magnetism in small compact clusters.

In the unrestricted Hartree-Fock (UHF) approximation the ground state 
for $\nu$ electrons is a single Slater determinant that can be  written as  
\begin{equation}
 |{\rm UHF}\rangle = \left[\;\prod_{k}^\nu \hat a_k^\dagger\;\right] 
                    |{\rm vac} \rangle \; .
\label{eq_fcnonda}
\end{equation}
The  single-particle states
\begin{equation}
 \hat a_k^\dagger = \sum_{l,\sigma = \uparrow, \downarrow} 
 A_{l\sigma}^k \hat c_{l\sigma}^\dagger
\label{eq_state}
\end{equation}
are linear combinations of the atomic-like orbitals associated to 
$\hat c_{l\sigma}^\dagger$. Notice that in Eq.~(\ref{eq_state}) we 
allow for the most general superposition of single-electron states 
since $\hat a_k^\dagger$ may involve a mixture of both $\uparrow$ 
and $\downarrow$ spin components. The eventually complex coefficients 
$ A_{i\sigma}^k $ are determined by minimizing the energy expectation 
value $E_{\rm UHF} = \langle {\rm UHF}| H|{\rm UHF}\rangle$. In
terms of the density matrix
\begin{equation} 
\rho_{l\sigma,m\sigma'} \equiv 
\langle {\rm UHF} |\hat c_{l\sigma}^\dagger\hat c_{m\sigma '}
|{\rm UHF}\rangle = 
\sum_{k=1}^{\nu} \bar A_{l\sigma}^k A_{m\sigma '}^k \;,
\label{eq_rho}
\end{equation}
this is given by
\begin{equation}
E_{\rm UHF}=-t\sum\limits_{\langle l,m \rangle , \sigma}\rho_{l\sigma ,m\sigma}
+U\sum_{l} \left(\rho_{l\uparrow,l\uparrow}\; \rho_{l\downarrow,l\downarrow}
-|\rho_{l\uparrow, l\downarrow}|^2\right) \; .
\label{eq_ener1}
\end{equation}
The energy minimization and the normalization constraints on 
the wave function lead to the usual self-consistent equations 
\begin{equation}
-t\sum_{m}A_{m\sigma}^k + U\left( A_{l\sigma}^k 
\rho_{l\bar\sigma,l\bar\sigma}
 - A_{l\bar\sigma}^k \rho_{l\bar\sigma,l\sigma}\right) = 
\varepsilon_k A_{l\sigma}^k \; .
\label{eq_hfe}
\end{equation}
For a given solution, the average  local electronic density 
$\langle n_{l} \rangle$ is given by
\begin{eqnarray}
 \langle n_{l} \rangle &=& \rho_{l\uparrow,l\uparrow}
 + \rho_{l\downarrow,l\downarrow} \; ,
\label{eq_ocup}
\end{eqnarray}
and the spin polarizations vectors 
$\langle \vec S_l \rangle = 
       (\langle S_l^x\rangle, \langle S_l^y\rangle, \langle S_l^z\rangle)$ 
by
\begin{eqnarray}
\langle  S_l^x\rangle &=&\left( \rho_{l\uparrow,l\downarrow}+
\rho_{l\downarrow,l\uparrow}\right)/2 \; , 
\label{eq_espinx}\nonumber\\
\langle  S_l^y\rangle &=&
-i\;\left( \rho_{l\uparrow,l\downarrow}-
\rho_{l\downarrow,l\uparrow}\right)/2 \; ,
\label{eq_espiny}\\
\langle  S_l^z\rangle &=&\
\left( \rho_{l\uparrow,l\uparrow}-
\rho_{l\downarrow,l\downarrow}\right)/2 \; .
\label{eq_espinz}\nonumber
\end{eqnarray}
The usual collinear UHF approach is recovered when 
$\rho_{l\sigma,l\bar\sigma}=0\;\;\forall \; l$, i.e., when all 
magnetic moments $\langle \vec S_l\rangle$ are parallel to $z$.
In practice, several random spin arrangements are used as starting 
points of the selfconsistent procedure in order to ensure that the 
final result corresponds to the true UHF ground state. In case of 
multiple selfconsistent solutions (nonequivalent by rotations) the 
UHF energies are compared. $E_{\rm UHF}$ can be rewritten as
\begin{equation}
E_{\rm UHF} =-t \sum_{\langle l, m \rangle , \sigma}\rho_{l\sigma,m\sigma}
+\frac{U}{4}\sum_{l}\langle n_{l} \rangle ^2
-U\sum_{l}|\langle \vec S_{l} \rangle|^2 \; .
\label{eq_ener2}
\end{equation}
One observes that the Hartree-Fock Coulomb energy $E_C^{HF}$
--- the sum of  the 2nd and 3rd terms in Eq.~(\ref{eq_ener2}) --- 
favors a uniform density distribution and the formation of local 
moments $\langle \vec S_l\rangle$. Due to the local character of 
Hubbard's Coulomb interaction, the relative orientation of the different 
$\langle \vec S_l\rangle$ does not affect $E_C^{HF}$. It is therefore 
the optimization of the kinetic energy that eventually leads to the 
formation of complex magnetic structures with 
$|\langle \vec S_l\rangle\!\cdot \!\langle \vec S_m\rangle|\not=1$
or to non-uniform density distributions $\langle n_l\rangle$.

As a result of the tendency to avoid double orbital occupancies, the UHF
solutions often correspond to states of broken symmetry: spin-density
waves (SDW's), charge density waves (CDW's) or both. The spin-rotational 
invariance of Eqs.~(\ref{eq_fcnonda}--\ref{eq_hfe}) implies that the 
energy is unchanged after a rotation of the whole spin arrangement
$\{\langle \vec S_l\rangle,\; l=1,\dots, N\}$. Therefore, if 
$\langle \vec S_l\rangle \not= 0$ one has a set of linearly independent 
congruent solutions $|{\rm UHF}k\rangle$ ($k\ge 2$) which have the same 
average energy $E_{\rm UHF}$ and which differ {}from each other only by 
the orientation of the spin polarizations relative to the cluster 
structure. The illustrations of spin arrangements shown in 
Sec.~\ref{sec:nocolin} correspond to one of these SDW's, which is chosen
only for the sake of clarity. The set of UHF solutions may be 
used to restore the symmetry appropriate to the exact ground state 
$|\Psi_{\rm 0}\rangle$ thereby improving the approximate wave function. 
For instance, for some SDW's having $\sum_l \langle \vec S_l\rangle=0$
one may consider the spin-symmetrized Hartree-Fock (SSHF) wave 
function \cite{fal2,stefan} 
\begin{equation} 
|{\rm SSHF}\rangle ={{ 
|{\rm UHF1}\rangle +
|{\rm UHF2}\rangle }\over \sqrt{2(1 +   
\langle{\rm UHF1}|{\rm UHF2}\rangle)}} \; , 
\label{eq_sshf}
\end{equation}
where $|{\rm UHF2}\rangle$ is obtained form $|{\rm UHF1}\rangle$
by interchanging up and down spins. In spite of its simplicity, 
$|{\rm SSHF}\rangle$ goes beyond the UHF approximation and corresponds 
to a correlated state satisfying
\begin{eqnarray} 
|\langle{\rm SSHF}|\Psi_{\rm 0}\rangle|& =& 
 \sqrt{2/(1 +   
\langle{\rm UHF1}|{\rm UHF2}\rangle)}\;|\langle{\rm UHF1}|\Psi_{\rm 0}\rangle|
\nonumber\\
&\ge&
|\langle{\rm UHF1}|\Psi_{\rm 0}\rangle|=
|\langle{\rm UHF2}|\Psi_{\rm 0}\rangle| \; .
\label{eq_sshf2}
\end{eqnarray} 
If quantum fluctuations between the two SDW's $|{\rm UHF1}\rangle$ and
$|{\rm UHF2}\rangle$ are non-negligible
($\langle{\rm UHF1}|H|{\rm UHF2}\rangle  \not= 
\langle{\rm UHF1}|{\rm UHF2}\rangle E_{\rm UHF}$) 
an energy reduction relative to $E_{\rm UHF}$ is obtained since
\begin{equation} 
E_{\rm SSHF} ={{ 
E_{\rm UHF} +
\langle{\rm UHF1}|H|{\rm UHF2}\rangle }\over {1 +   
\langle{\rm UHF1}|{\rm UHF2}\rangle}} \; .  
\label{eq_sshf3}
\end{equation}
$\Delta E=E_{\rm UHF}-E_{\rm SSHF}$ measures the importance of these
quantum spin fluctuations. More complex symmetrized states involving a
linear combination of several degenerate UHF states may be constructed 
analogously.

In order to quantify the role of electron correlations and to assess 
goal and limitations of the UHF approximation in applications to finite 
clusters, we shall also compare some of our results with those obtained 
by applying exact diagonalization methods \cite{exact2,prlgus,prbflo}.
In this case the Hubbard model is solved numerically by expanding 
its ground-state $\vert\Psi_0\rangle$ in a complete set of basis states 
$\vert\Phi_m\rangle$ which have definite occupation numbers $n^m_{l \sigma}$ 
at all orbitals $l \sigma$ 
($\hat n_{l\sigma}   \vert\Phi_m\rangle = n_{l\sigma}^m \vert\Phi_m\rangle$, 
with $n_{l\sigma}^m = 0$, $1$). $\vert\Psi_0\rangle$ is written as
\begin{equation}
\vert\Psi_0\rangle = \sum_m \alpha_{lm}  \vert\Phi_m\rangle \; ,
\end{equation}
where 
\begin{equation}
\label{eq:VBexact}
\vert\Phi_m\rangle = \left[\prod_{l\sigma} 
(\hat c^\dagger_{l\sigma})^{n^m_{l \sigma}}
\right] \vert vac\rangle\; .
\end{equation}
The values of $n^m_{l \sigma}$ satisfy the usual conservation of the number 
of electrons $\nu  = \nu_\uparrow + \nu_\downarrow$ and of the $z$ component 
of the total spin $S_z = (\nu_\uparrow - \nu_\downarrow)/2$, where
$\nu_\sigma = \sum_{l} n^m_{l	 \sigma}$. Taking into account all
possible electronic configurations may imply a considerable numerical 
effort which depends on the number of atoms and on band filling.
For not too large clusters, the expansion coefficients $\alpha_{lm}$ 
can be determined by sparse-matrix diagonalization procedures. 
The results  presented in this work were obtained using a Lanczos 
iterative method \cite{prlgus,prbflo,lan}. In order to calculate 
$\vert\Psi_0\rangle$ one usually works in the subspace 
of minimal $S_z$ since this ensures that there are no {\em a priori\/} 
restrictions on the total spin $S$. The ground-state spin $S$ is then 
obtained by applying to $\vert\Psi_0\rangle$ the total spin operator
\begin{equation}
\hat S^2 = \sum_{lm} {\vec S}_l\cdot {\vec S}_m = 
\sum_{lm} \left[ {1\over 2} (\hat S_l^+ \hat S_m^- + 
                             \hat S_l^- \hat S_m^+ )  
                           + \hat S_l^z \hat S_m^z\right]\; ,
\end{equation}
where 
$\hat S_l^+ = \hat c^\dagger_{l\uparrow} \hat c_{l\downarrow}$, 
$\hat S_l^- = \hat c^\dagger_{l\downarrow} \hat c_{l\uparrow}$ and
$\hat S_l^z = (\hat n_{l\uparrow} - \hat n_{l\downarrow})/2$.
{}From the expectation values of $\vec S_l\!\cdot \!\vec S_m$ one
also obtains the local magnetic moments ($l=m$) and the intersite
spin correlation functions ($l \not= m$).

\section{Noncollinear magnetic order in compact clusters}
\label{sec:nocolin}

In Fig.~\ref{fig:mom_fcc} UHF results are given for the local magnetic
moments $\mu_l=|\langle \vec S_l \rangle|$ and the total magnetic moment 
$\mu_T=|\sum_l \langle \vec S_l \rangle|$ of fcc-like clusters having
$N=13$, $19$ and $43$ atoms at half band filling ($\nu=N$). These 
clusters are formed by adding to a central atom ($l=1$) the successive
shells of its first NN's ($N=13$), second NN's ($N=19$) and third
NN's ($N=43$). Several common properties are observed as a function of 
$U/t$. Starting {}from the uncorrelated limit ($U=0$), the total moment 
$\mu_T$ remains approximately constant for $U/t\le 3$--$4$. In the 
weakly interacting regime, the local moments $\mu_l$ do not depend 
strongly on $U/t$ and are generally small. Notice that for $N=43$, 
$\mu_T$ is not minimal at small $U$ ($\mu_T = 3/2$) due to degeneracies 
in the single-particle spectrum. For larger $U/t$, $\mu_T$ decreases 
rapidly eventually with some discontinuities and a few oscillations 
reaching values close to $\mu_T =0$ for $U/t\simeq 5$--$6$ 
($U/t\simeq 9$ for $N=43$). At this intermediate range of $U/t$, the 
$\mu_l$ increase more rapidly reaching values not far {}from saturation 
at the $U/t$ for which $\mu_T$ is minimum ($\mu_l \simeq 0.40$--$0.45$). 
The opposite trends shown by $\mu_l$ and $\mu_T$ are a clear indication 
of the expected onset of strong AF-like order at half-band filling. 
We shall see in the following that this corresponds in fact to 
noncollinear spin arrangements. If $U/t$ is  increased beyond $U/t=5$--$6$ 
(beyond $U/t=9$ for $N=43$) the local moments do not vary significantly, 
and $\mu_T$  either does not change very much ($N=19$) or increases 
monotonically ($N=13$ and $43$) remaining always smaller than $1/2$. 

The changes in $\mu_l$ and $\mu_T$ are the result of qualitative changes 
in the magnetic order. As an example we show in Fig.~\ref{fig:stfcc13} 
the selfconsistent spin arrangements obtained in fcc clusters with 
$N=13$ atoms for representative values of $U/t$ 
($\nu = N$) \cite{foot_sim}. 
For small $U/t$ ($U/t<3.7$) one finds a collinear AF order with small 
$\mu_l$ [Fig.~\ref{fig:stfcc13}(a)]. Here we observe a charge and
spin density-wave at the cluster surface  that is related
to  degeneracies in the single-particle spectrum at $U=0$. 
The atoms belonging to the central (001) plane [shown in grey in 
Fig.~\ref{fig:stfcc13}(a)] have much larger moments than the atoms 
at the upper and lower (001) planes. For example for $U/t=0.5$, 
$\mu_l = 0.26{\mu_{\rm B}}$ at the the central plane, while the other 
surface moments are $\mu_l = 0.0046{\mu_{\rm B}}$ 
(see Fig.~\ref{fig:mom_fcc}). The central atom has a small spin 
polarization ($\mu_1 =0.002  \mu_{\rm B}$ for $U/t = 0.5$).   
Notice that the magnetic order within the upper and lower (001) planes 
is ferromagnetic-like (with small $\mu_l$) and that 
the surface moments belonging to successive  (001) planes couple 
antiferromagnetically. Thus, the magnetic moments at the surface of the 
central plane are not frustrated since all its NN's are antiparallel to 
them. This explains qualitatively the larger $\mu_l$ found at this plane.
In contrast, unavoidable frustrations are found for the
smallest magnetic moments at the central site and at the atoms
of the upper and lower (001) planes [see Fig.~\ref{fig:stfcc13}(a)]

The crossover {}from the small-$U/t$ to the large-$U/t$ regime takes 
place as a succession of noncollinear spin arrangements which attempt 
to minimize the magnetic frustrations among the increasing local moments. 
A representative example is shown in Fig.~\ref{fig:stfcc13}(b). 
While in this case the spin arrangement is noncollinear, all the spin 
moments still lie in the same plane. The very small values of $\mu_T$ 
shown in Fig.~\ref{fig:mom_fcc} for intermediate $U/t$ indicate that there 
is an almost complete cancellation among the  $\langle \vec S_l\rangle$.
However, this type of spin arrangements are quite unstable
if the strength of Coulomb interactions is further increased. 
At $U/t\simeq 5.1$ the cluster adopts a fully 3-dimensional spin 
structure which remains essentially unchanged even in the strongly 
correlated limit [Fig.~\ref{fig:stfcc13}(c)]. 
The spin arrangement can be viewed as 
a slight distortion of the  spin ordering that minimizes the energy 
of a classical AF Heisenberg model on the  surface shell, i.e., 
ignoring the interactions with the central site. In fact, if the central 
atom were removed or if it carried no local moment, as it is 
the case for $\nu=12$, the surface moments  $\langle \vec S_l\rangle$ 
would point along the medians of one of the triangles at the surface
and would lie all within a plane. The magnetic interactions with the 
central spin $\langle \vec S_1\rangle$ in the 13-atom cluster induce 
a small tilt $\langle S_l^z\rangle$ of the surface spin polarizations 
$\langle \vec S_l\rangle$ which is opposite to $\langle \vec S_1\rangle$
($\langle \vec S_1\rangle \parallel \hat z$ for $\nu=N=13$). 
The $\langle S_l^z\rangle$ component is the same for all surface sites 
and depends moderately on $U/t$ ($|\langle S_l^z\rangle| = 0.023$--$0.056$ 
for $U/t\ge 5.1$).

Similar magnetic structures are also found in larger symmetric fcc 
clusters. In Fig.~\ref{fig:stfcc43} the self-consistent spin 
configuration for $N=43$ and $U/t=10$ is illustrated ($\nu=N$).
The moment $\langle \vec S_1\rangle$ of the central site points 
along the (111) direction ($\perp$ to the plane of Fig.~\ref{fig:stfcc43}) 
\cite{foot_sim}. The other $\langle \vec S_m\rangle$ lie almost in the
plane of the figure with  only small components $\langle S^z_m\rangle$ 
along the (111) direction. As for $N=13$, $\langle S_m^z\rangle$ is
induced by the interactions with $\langle \vec S_1\rangle$. In fact, if 
$\langle \vec S_1\rangle$ vanished, all the spins would be
in the plane of the figure. $\langle S^z_m\rangle$ is the same for all 
atoms in a given NN shell, and changes sign as we move {}from the center 
to the  surface of the cluster. For example for $U/t=10$, 
$\cos\theta_{1m}=-0.22$ for $m$ belonging to the first shell,  
$\cos\theta_{1m}=0.25$ for $m$ in the second shell and 
$\cos\theta_{1m}=0.01$ for $m$ in the third shell 
($\cos\theta_{lm} = \langle \vec S_l \rangle \cdot
\langle \vec S_m \rangle/ |\langle \vec S_l\rangle||\langle \vec S_m\rangle|$).
Notice that the spin ordering of the innermost 13 atoms is similar 
to the one found in fcc clusters with $N=13$ [Fig.~\ref{fig:stfcc13}(c)]. 
While these trends hold for symmetric fcc clusters, it is worth to 
remark that strong modifications of the magnetic order generally occur
if the symmetry of the cluster is lowered. For instance, for $N=14$ 
--- an atom added to the closed-shell 13-atom cluster --- 
one obtains a ferromagnetic-like coupling between the central atom and 
some of the first neighbors ($\nu=N$ and $U/t=10$).

A more detailed account of the spin correlations in the UHF ground state 
of fcc clusters is provided by the expectation values 
$\langle \vec S_l\!\cdot\!\vec S_m\rangle$. In Fig.~\ref{fig:spcf13-43} 
the average of $\langle \vec S_l\!\cdot\!\vec S_m\rangle$ between NN 
atoms at different shells is shown. In most cases we observe that 
$\langle \vec S_l\!\cdot\!\vec S_m\rangle < 0$ (AF correlations) and 
that $|\langle \vec S_l\!\cdot\!\vec S_m\rangle|$ increases with 
increasing $U/t$. A particularly important increase of AF correlations 
is observed for intermediate $U/t$ when the local moments $\mu_l$ are 
formed ($U/t\simeq 4.5$--$5.5$ for $N=13$ and $19$, see also 
Fig.~\ref{fig:mom_fcc}). This is consistent with the already discussed 
decrease of $\mu_T$ and the onset of AF order. There are, however, 
two exceptions to this trend. The correlations $\gamma_{01}$ between 
the central site and the surface shell for $N=13$ increase somewhat 
for $U/t \simeq 5$ ($\gamma_{01}<0$) implying that these AF correlations 
first tend to be less important when the $\mu_l$ increase. 
It seems that, as a result of frustrations, the more important increase 
of AF correlations $\gamma_{11}$ among the surface NN's is done at 
the expense of the correlations between the central atom and the surface. 
A similar behavior is found in exact calculations, as it will be shown 
in Sec.~\ref{sec:corr}. Another interesting case concerns the correlations 
$\gamma_{12}$ between the first and second atomic shells for $N=19$ and 
$43$. Here we observe that $\gamma_{12}$ changes sign upon going {}from 
$N=19$ to $N=43$ ($U/t>5$). Moreover, once the local moments are formed, 
$\gamma_{01}$ decreases for $N=19$ [$\gamma_{01}(19) < 0$] 
and increases for $N=43$ as $U/t$ increases 
[$\gamma_{01}(43) > 0$]. This behavior can be qualitatively understood by
comparing the surfaces of these clusters. For $N=19$ the outermost spins 
(shell 2) are free to couple antiferromagnetically with their NN's 
(shell 1). However, the presence of an additional atomic shell for $N=43$, 
which has NN's belonging to both the first and second shells, forces a 
parallel alignment of the spins in shells 1 and 2 in order to allow AF 
coupling with the third shell. In fact, as shown in 
Fig.~\ref{fig:spcf13-43}(c), $\gamma_{13}$ and $\gamma_{23}$ 
are the strongest AF correlations in the 43-atom cluster. The same 
conclusions are drawn by comparing the relative orientations of the spin 
polarizations (Fig.~\ref{fig:stfcc43}). In particular we observe that 
the angles $\theta_{23}$ between NN spin polarizations at shells 2 and 3 
are the largest in average ($\theta_{23} = \pi$ or close to it).

\begin{figure}
\begin{center}
\leavevmode
\epsfxsize=8.50cm
\epsfbox{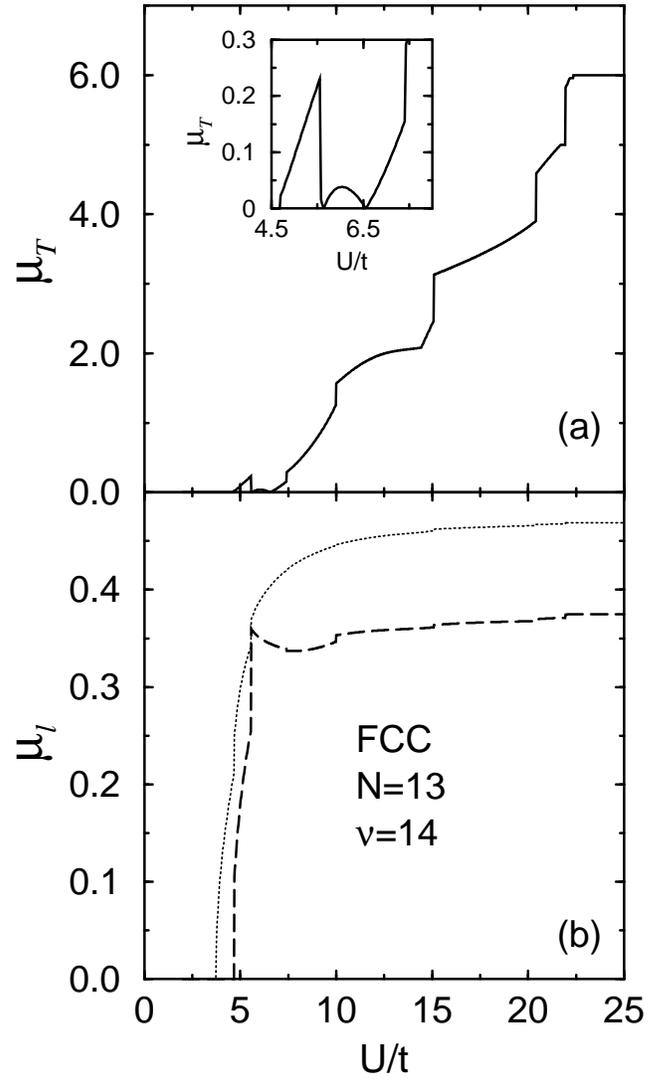}
\end{center}
\caption{
Magnetic moments in an fcc-like 13-atom cluster with $\nu=14$ electrons
as a function of $U/t$ (UHF approximation): (a) total moment 
$\mu_T=|\sum_l \langle \vec S_l\rangle |$ and 
(b) local moments $\mu_l=|\langle \vec S_l\rangle |$ at
the central site (dashed line) and at the surface (dotted line).
}
\label{fig:mom_14}
\end{figure}

In addition to the AF half-filled case it is also interesting to discuss 
other band fillings, for example $\nu=N+1$ that is known to develop a 
FM ground state in the limit of large $U/t$ \cite{nag}. In 
Fig.~\ref{fig:mom_14} results are given for $\mu_l$ and $ \mu_T$ in an 
fcc 13-atom cluster with $\nu=14$ electrons. In this case we observe 
an essentially monotonic increase of $\mu_T$ that reflects the 
progressive development of a fully polarized ferromagnetic state.  Close
to the threshold for the onset of ferromagnetism ($4.5\le U/t \le 6.5$) 
the changes of the AF-like spin arrangement produce small oscillations 
of $\mu_T$ (see Fig~\ref{fig:mom_14}).
The approximately step-like behavior resembles at first sight
the results obtained in collinear mean-field calculations\cite{hugo}. 
However, in the present case the physical picture behind the formation 
of a FM state is quite different. The increase of $\mu_T$ close to the steps
involves a succession of noncollinear spin arrangements with increasing 
degree of parallel moment alignment. Moreover, notice that  the local 
moments increase much more rapidly than $\mu_T$ approaching saturation 
($\mu_l\simeq 0.40$--$0.45$) for values of $U/t$ at which $\mu_T$ is
still small. Thus, the increase of $\mu_T$ with $U/t$ is the
result of the parallel alignment of already existing local moments 
$\mu_l$. In contrast, in collinear Hartree-Fock calculations the increase 
of $\mu_T$ is associated to the formation of the local moments themselves
since $\mu_T$ approximately proportional to $\mu_l$ \cite{hugo}. 
Comparison with exact calculations shows that the present noncollinear 
picture is qualitatively closer to the actual ground-state magnetic properties 
than the one derived in collinear calculations, at least for the single-band 
Hubbard model. Indeed, we have computed the local moments 
$\mu_l^2= \langle S^2_l \rangle$ and the total spin $S$ of the same 
fcc 13-atom cluster ($\nu=14$) as a function of $U/t$ by using a Lanczos 
exact diagonalization method. Starting {}from the uncorrelated limit
[$\mu^2_1(U\!=\!0) = 0.28$ and   $\mu^2_l(U\!=\!0)  = 0.36$ for $l=2$--$13$)]
 $\mu^2_l $  increases rapidly with $U/t$ reaching values close to saturation 
already for $U/t \simeq 10$ [$\mu^2_1(U\!=\!10)  = 0.56$ and 
$\mu^2_l(U\!=\!10)  = 0.64$ while $\mu^2_1(U\!=\!\infty) = 0.56$ and 
$\mu^2_l(U\!=\!\infty)  = 0.70$ ($l=2$--$13$)]. In contrast the ground-state 
spin $S$ remains equal to zero up to $U/t \simeq 40$. This implies that for 
$\nu = N+1= 14$ the local moments are formed well before FM order sets in, 
as observed in the noncollinear UHF calculations (Fig.~\ref{fig:mom_14}). 
Notice, however, that the situation could be different for other band 
fillings where ground-state ferromagnetism is found at much 
smaller $U/t$ or where $S$ is a non-monotonous function of $U/t$ 
(e.g., for $\nu=15$--$18$ \cite{prbflo}). UHF yields larger local magnetic 
moments at the cluster surface in agreement with the exact results but 
it underestimates severely both the Coulomb repulsion $U/t$ above which 
$\mu_T>1/2$ and the $U/t$ for reaching saturation.

Calculations have been also performed for fcc-like 13-atom clusters having 
$\nu = N-1 = 12$ electrons. For small $U$ ($U/t<4.9$) the calculated 
magnetic order is collinear ($\langle \vec S_l\rangle \parallel \hat z$). 
The UHF ground state is a broken symmetry state that shows a
charge- and spin-density wave along the (001) direction. For $U/t < 3.5$ 
the atoms $l$ in the central (001) plane present an enhanced electron 
density $\langle n_l\rangle \simeq 1.12$ and local magnetic moments 
$\mu_l=0$. The atoms of the upper and lower (001) planes show 
AF order within each plane and $\mu_l=0.17$. The vanishing $\mu_l$ at 
the central (001) plane can be interpreted as a consequence of magnetic 
frustrations since the sum of their NN spins is zero. For $U/t> 3.5$, 
the surface atoms at the central (001) plane develop local moments 
$\mu_l\not= 0$. The spin arrangement remains collinear with important 
frustrations at some triangular faces. For $U/t > 4.9$ the spin-density 
distribution changes to noncollinear with all surface spin having the 
same modulus. However, the central site remains unpolarized. In fact 
the magnetic order is the same as the one obtained if the central site 
is removed ($N=12$). The moments at the triangular faces point along 
the medians just as in the  classical antiferromagnetic ground state 
of an isolated triangle\cite{maol}. In spite of the three-dimensional 
geometry of the cluster the spin structure is two-dimensional 
(all magnetic moments lie on the same plane).

The topology and symmetry of the cluster structure plays a major
role in determining the magnetic order and magnetic correlations 
within the cluster. In this context, it is interesting to consider 
different geometries and to compare their magnetic behavior. 
In Fig.~\ref{fig:stico13} the UHF magnetic 
order in a 13-atom icosahedral cluster having $\nu=12$ electrons is 
illustrated for representative values of $U/t$. As a result of 
degeneracies in the single-particle spectrum ($U=0$)
a noncollinear, coplanar arrangement of the local magnetic moments 
is found at the surface already for very small $U/t$ 
($U/t<5$, see Fig.~\ref{fig:stico13}). At the central atom (not shown 
in the figure) the magnetic moment $\mu_1=0$. A similar behavior is 
observed in fcc clusters ($N=13$, $\nu=12$). The tendency to avoid 
frustrations within NN triangular rings conditions 
the arrangement of the local moments. In some triangles 
the spin polarizations point along the medians just as in the classical 
Heisenberg model, while in others two spins are parallel (i.e., fully 
frustrated). For larger values of $U/t$ ($U/t>5$) a less frustrated 
solution is favored by the antiferromagnetic correlations. This 
corresponds to a truly three-dimensional arrangement of the surface 
moments with $\mu_1=0$. The spin structure is such that if
the surface $\langle \vec S_l\rangle$ are brought to a common origin
they form an icosahedron. The magnetic moments $\langle \vec S_l\rangle$ 
on pentagonal rings present a small component that is perpendicular to the 
plane containing the atoms and that is antiparallel to the magnetic moment 
of the atom capping the ring. The projections of $\langle \vec S_l\rangle$ 
on to the plane of the ring are ordered in the same way as in an  
isolated pentagon.

\section {Comparison between UHF and exact results}
\label{sec:corr}

\begin{figure}
\begin{center}
\leavevmode
\epsfxsize=8.5cm
\epsfbox{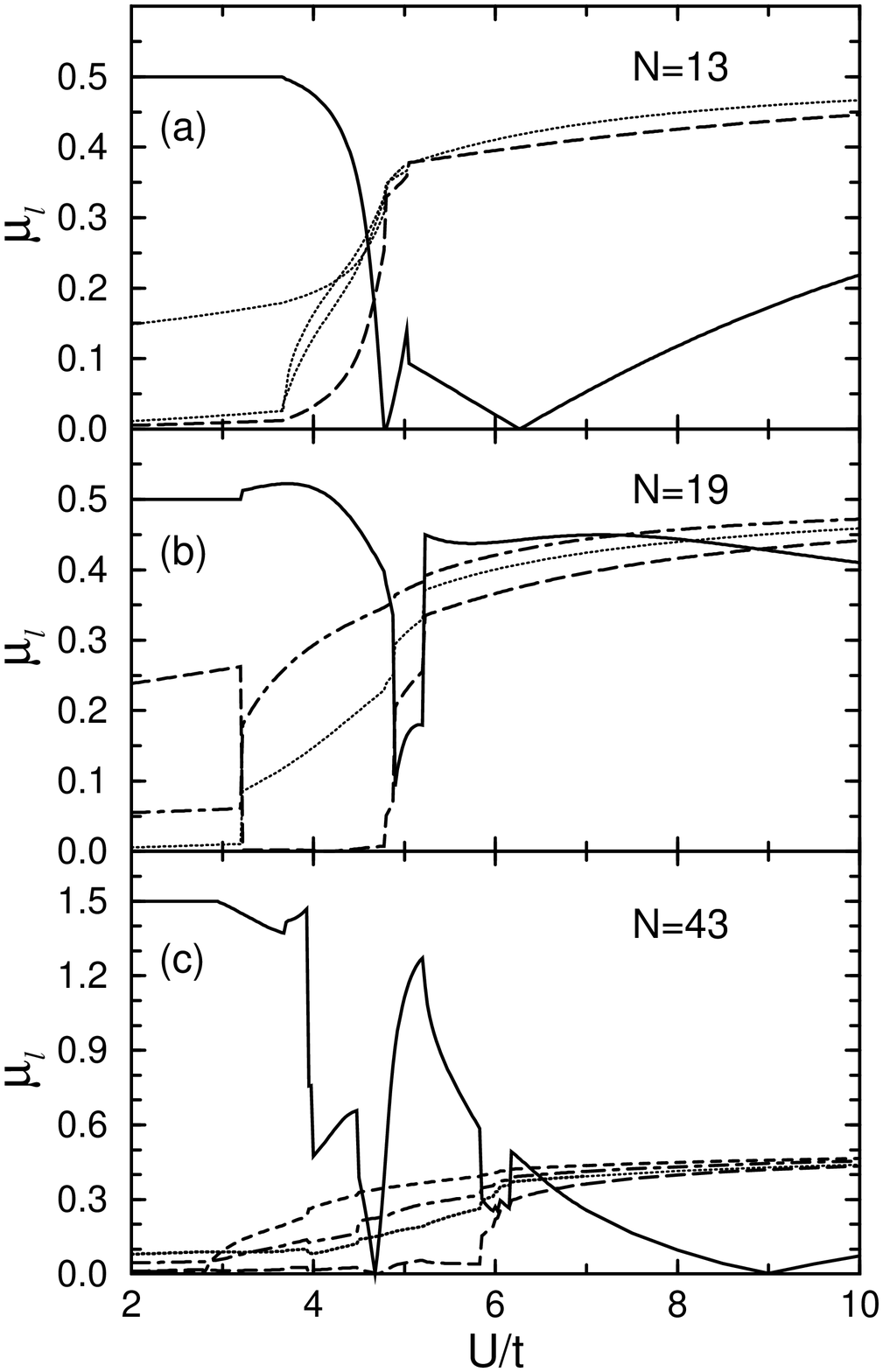}
\end{center}
\caption{
Magnetic moments as a function of the Coulomb repulsion strength $U/t$
for  fcc-like clusters having (a) $N= 13$, (b) $N= 19$  and (c) $N= 43$ 
atoms at half-band filling (UHF approximation).
Local moments $\mu_l=|\langle \vec S_l \rangle|$ are shown for the 
central atom $l=1$ (dashed), its first NN's ($l=2$--$13$, 
dotted), second  NN's ($l=14$--$19$, dashed-dotted)
and third NN's ($l=20$--$43$, short dashed). In (b) and (c)
$\mu_l$ refers to the shell average.
The  total moment $ \mu_T = |\sum_l \langle \vec S_l \rangle|$ is 
given by the solid curves.
}
\label{fig:mom_fcc}
\end{figure}

\begin{figure}
\begin{center}
\leavevmode
\epsfxsize=8.5cm
\epsfbox{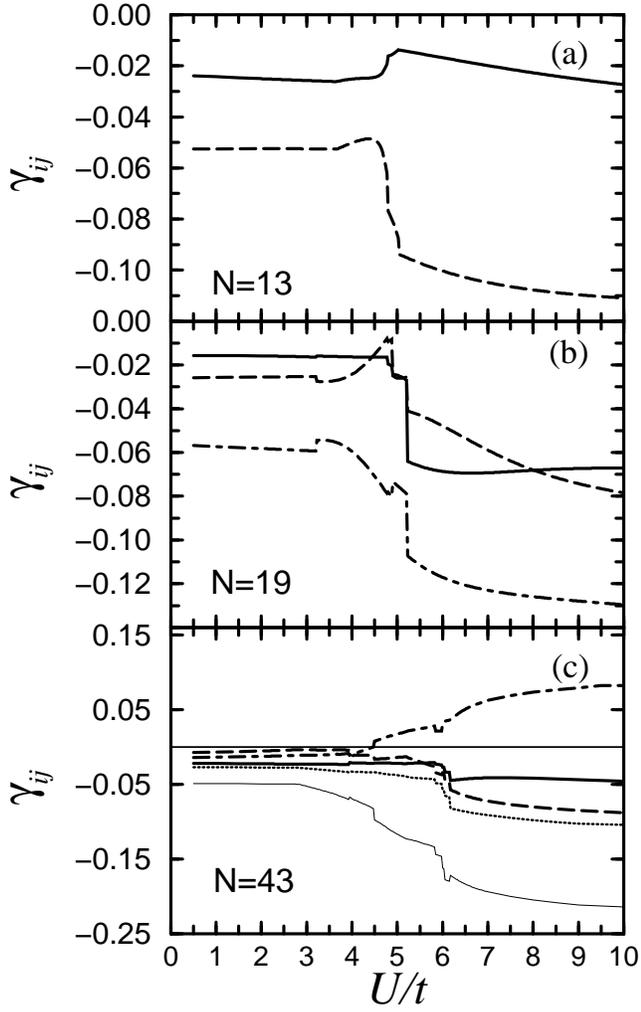}
\end{center}
\caption{
Spin correlations $\langle \vec S_l \cdot \vec S_m\rangle$ 
in fcc-like clusters having $N=13$--$43$ atoms as a function of $U/t$. 
Results are given for the average spin correlations $\gamma_{01}$ between the 
central site and its first NN's (solid), $\gamma_{11}$ between NN 
sites on the first shell (dashed), $\gamma_{12}$ between NN sites on the 
first and second shell (dashed-dotted), $\gamma_{13}$ between NN sites on the 
first and third shell (dotted), and $\gamma_{23}$ between NN sites on the 
second and third shell [lower thin curve in (c)].
}
\label{fig:spcf13-43}
\end{figure}

\begin{figure}
\begin{center}
\leavevmode
\epsfxsize=8.50cm
\epsfbox{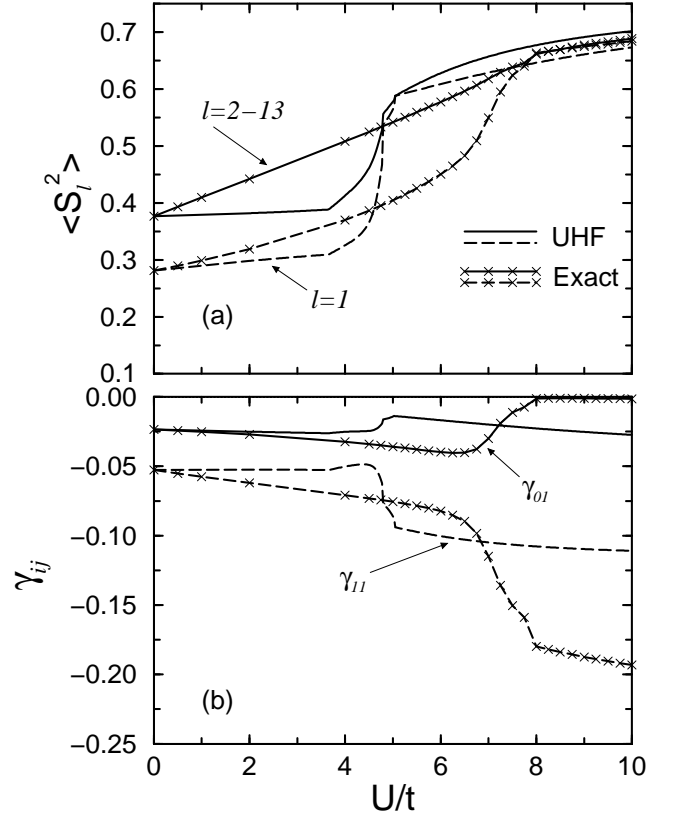}
\end{center}
\caption{ 
Comparison between UHF and exact results for an fcc-like 13-atom cluster
at half-band filling. (a) Local magnetic moments 
$\mu_l^2=|\langle \vec S_l\!\cdot\! \vec S_l\rangle|$
at the central site ($l=1$, dashed curves) and at the cluster surface 
($l=2$--$13$, solid curves). (b) Average $\gamma_{ij}$ of the spin 
correlation functions $\langle \vec S_l\!\cdot\! \vec S_m\rangle$ between 
the central site and its first NN's at the surface ($\gamma_{01}$, 
dashed curves) and between NN's at the surface shell ($\gamma_{11}$, 
solid curves). The curves with (without) crosses correspond to exact 
(UHF) results.
        }
\label{fig:spc_fcc13}
\end{figure}

In Fig.~\ref{fig:spc_fcc13} UHF and exact results are given for the 
local moments $\mu_l^2=\langle S_l^2\rangle$ and spin correlations 
$\langle \vec S_l\!\cdot\!\vec S_m\rangle$ in an fcc-like 13-atom 
cluster. The UHF results for $\mu_l^2$ are quantitatively not far 
{}from the exact results. Not only the uncorrelated limit ($U=0$) is 
reproduced, but good agreement is also obtained in the large $U/t$ 
regime. Main trends such as the larger $\mu_l^2$ at the cluster 
surface for small $U/t$ and the reduction of the difference between 
surface and inner moments for large $U/t$ are correctly given. 
However, UHF underestimates the increase of $\mu_l^2$ 
for $U/t<3.7$ and anticipates the tendency to localization with 
increasing $U/t$. This results in a much more rapid crossover {}from 
weak to strong interacting regimes than in the exact solution. 
The quantitative differences are more important in the case of
spin correlation functions $\langle \vec S_l\!\cdot\!\vec S_m\rangle$
particularly for large $U/t$ [Fig.~\ref{fig:spc_fcc13}(b)]. Here we 
find that UHF underestimates the strength of AF spin correlations 
$\gamma_{11}$ at the surface, since the formation of permanent local 
moments blocks quantum spin fluctuations along the transversal directions. 
Still, in both UHF and exact calculations, the increase of AF
correlations at the surface (increase of $|\gamma_{11}|$) is
done at the expense of a decrease of the spin correlations with
the central atom (decrease of $|\gamma_{01}|$). The discrepancies 
between UHF and exact results for $N=13$ show the limits of mean-field
and give us an approximate idea of the corrections to be expected 
in correlated calculations on larger clusters. As expected, UHF yields 
better results for properties like the local moments, 
that are related to the density distribution, than for the correlation 
functions. Nevertheless, since the trends given by UHF
are qualitatively correct, it is reasonable to expect that
the conclusions derived for larger clusters are also valid 
(Figs.~\ref{fig:mom_fcc} and \ref{fig:spcf13-43}). 

\begin{figure}
\begin{center}
\leavevmode
\epsfxsize=5.7cm
\epsfbox{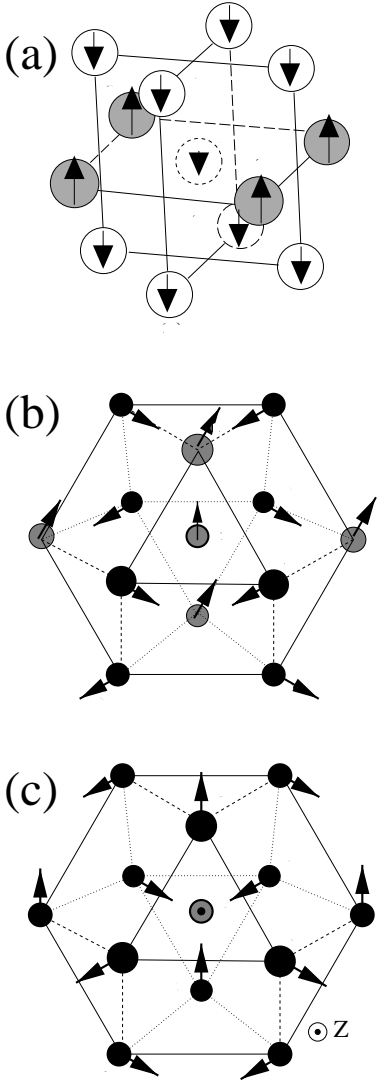}
\end{center}
\caption{
Illustration of the UHF magnetic order in  fcc-like clusters having 
$N=13$ atoms, $\nu=N=13$ electrons and  representative values of $U/t$
(Hubbard model) \protect\cite{foot_sim}. 
(a) $U/t < 3.7$:   
The surface atoms in grey have larger local magnetic moments
$|\langle \vec S_l\rangle| = 0.16$--$0.18$ and less electron density 
$\langle n_l\rangle\simeq 0.88$--$0.90$ than the other surface atoms 
for which $|\langle \vec S_l\rangle| \simeq 0.02$--$0.03$ and 
$\langle n_l\rangle=1.05$--$1.12$. The spin arrangement is collinear.
(b) $4.8\le U/t < 5.1$: 
The plane of the figure is perpendicular to the (111) direction. The 
arrows show the local spin polarizations $\langle \vec S_l\rangle$ 
which are all within the (111) plane. 
(c) $U/t \ge 5.1$: 
The plane of the figure is perpendicular to the (111) direction.
The arrows show the projection of $\langle \vec S_l\rangle$ on to 
the (111) plane. At the central atom $\langle \vec S_1\rangle\parallel (111)$.
The off-plane components of $\langle \vec S_l\rangle$ at the surface 
atoms are antiparallel to the central spin and have all the same value 
($|\langle S_l^z\rangle| \simeq 0.023$--$0.056$ for $l=2$--$13$). 
        }
\label{fig:stfcc13}
\end{figure}

\begin{figure}
\begin{center}
\leavevmode
\epsfxsize=8.cm
\epsfbox{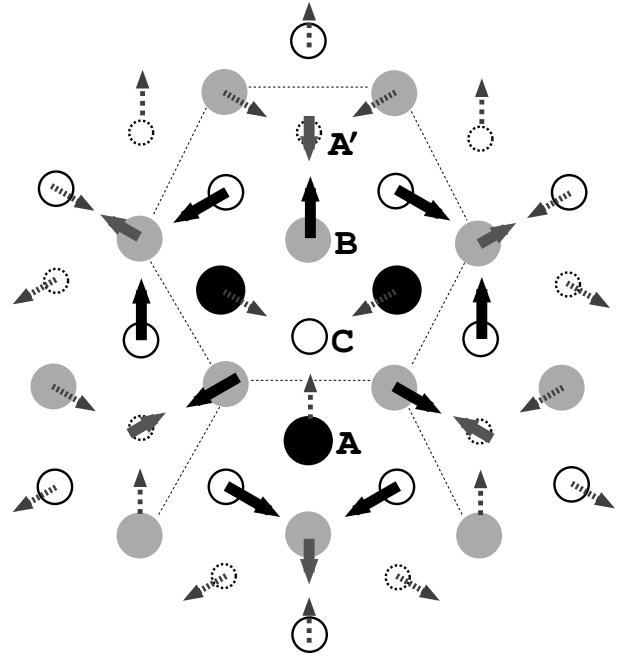}
\end{center}
\caption{
Illustration of the local magnetic moments $\langle \vec S_l\rangle$
in a 43-atom fcc-like cluster at half-band filling and $U/t = 10$
\protect\cite{foot_sim}. As in Fig.~\protect\ref{fig:stfcc13}(c), 
the plane of the figure is perpendicular to the (111) direction. 
The atoms are represented by circles using different sizes and
grey tones for different $(111)$ layers. The later are denoted by 
A, B, C and A' starting {}from above. The arrows show the projection 
of $\langle \vec S_l\rangle$ on to the (111) plane. At the central 
atom $\langle \vec S_1\rangle\parallel (111)$.
The spin structure is 3-dimensional with off-plane components of 
$\langle \vec S_l\rangle$ that alternate sign on different shells
around the central atom. For the 13 innermost atoms the magnetic
order is similar to the one shown in Fig.~\protect\ref{fig:stfcc13}(c).
        }
\label{fig:stfcc43}
\end{figure}

\begin{figure}
\begin{center}
\leavevmode
\epsfxsize=7.0cm
\epsfbox{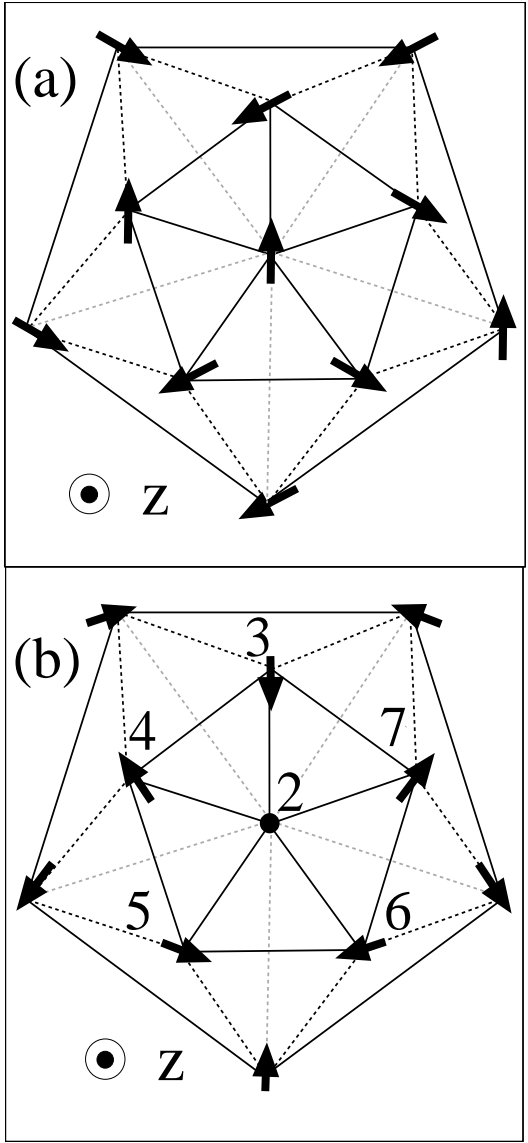}
\end{center}
\caption{
Illustration of the magnetic order at the surface of an icosahedral 
13-atom cluster with $\nu = N-1=12$ electrons (UHF approximation)
\protect\cite{foot_sim}.
(a) $0.8 < U/t \leq 5\;$: 
All local magnetic moments $\langle \vec S_l\rangle$, represented by 
the arrows, are parallel to the plane of the figure. 
(b) $U/t > 5\;$:
The arrows show the projections of $\langle \vec S_l\rangle$ on the 
plane of the figure. At site $l=2$, $\langle \vec S_l\rangle$ is 
perpendicular to the plane of the figure (along the $z$ axis). 
The $\langle \vec S_l\rangle$ at sites $l=3$--$7$ have antiparallel 
projection to $\langle \vec S_2\rangle$ 
[$cos \theta_{2l} = (\langle \vec S_l \rangle \cdot \langle \vec S_2\rangle) / 
(|\langle \vec S_l\rangle| |\langle \vec S_2\rangle|\simeq -0.45)$].
If brought to a common origin, the spin polarizations 
$\langle \vec S_l\rangle$ form a perfect icosahedron. Notice that 
the positions of the atoms in the lower pentagonal rings have been 
expanded along the polar radius in order to ease the visualization.
        }
\label{fig:stico13}
\end{figure}

\begin{figure}[x]
\begin{center}
\leavevmode
\epsfxsize=8.5cm
\epsfbox{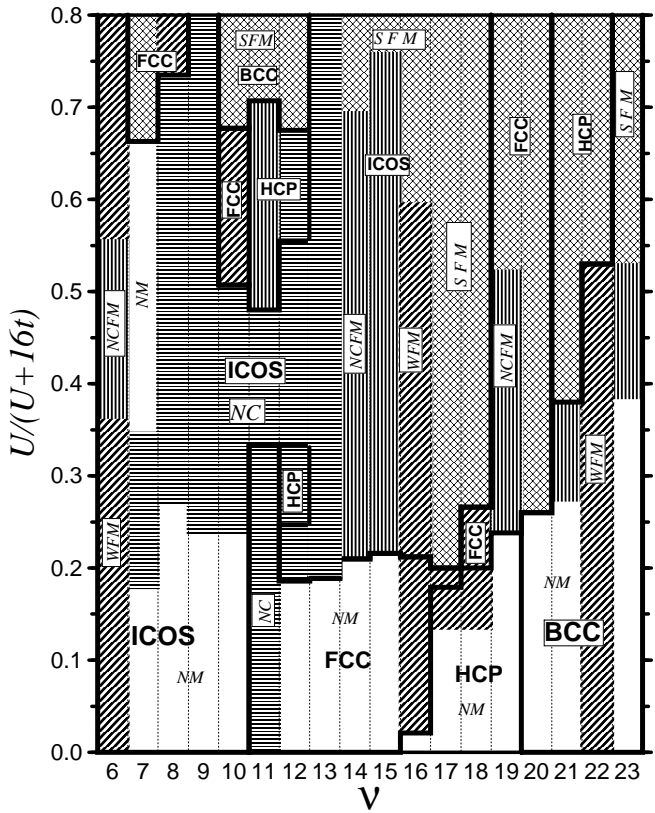}
\end{center}
\caption{
Magnetic and structural diagram of clusters having $N=13$  atoms as
obtained by using the Hubbard model in the unrestricted Hartree-Fock 
(UHF) approximation. Four different types of structures are considered:
icosahedral (ICOS), face centered cubic (FCC), hexagonal close packed 
(HCP), and body centered cubic (BCC). The lowest-energy structure is
given as a function of the Coulomb repulsion strength $U$, hopping 
integral $t$ and number of electrons $\nu$. The bold lines indicate the 
structural transitions. For $\nu < 6$ the lowest-energy structure is 
icosahedral and for $\nu>23$ it is bcc. The corresponding UHF magnetic 
orders are indicated by different shading patterns:
saturated ferromagnetic (crossed lines, {\em SFM}), 
collinear weak ferromagnetic (diagonal lines, {\em WFM}),
collinear nonmagnetic (no shading, {\em NM}),
noncollinear with total moment $\mu_T<1$ (horizontal lines, {\em NC}), and
noncollinear with total moment $\mu_T\ge 1$ (vertical lines, {\em NCFM}).
        }
\label{phdi_13_UHF}
\end{figure}

\begin{figure}[x]
\begin{center}
\leavevmode
\epsfxsize=6.5cm
\epsfbox{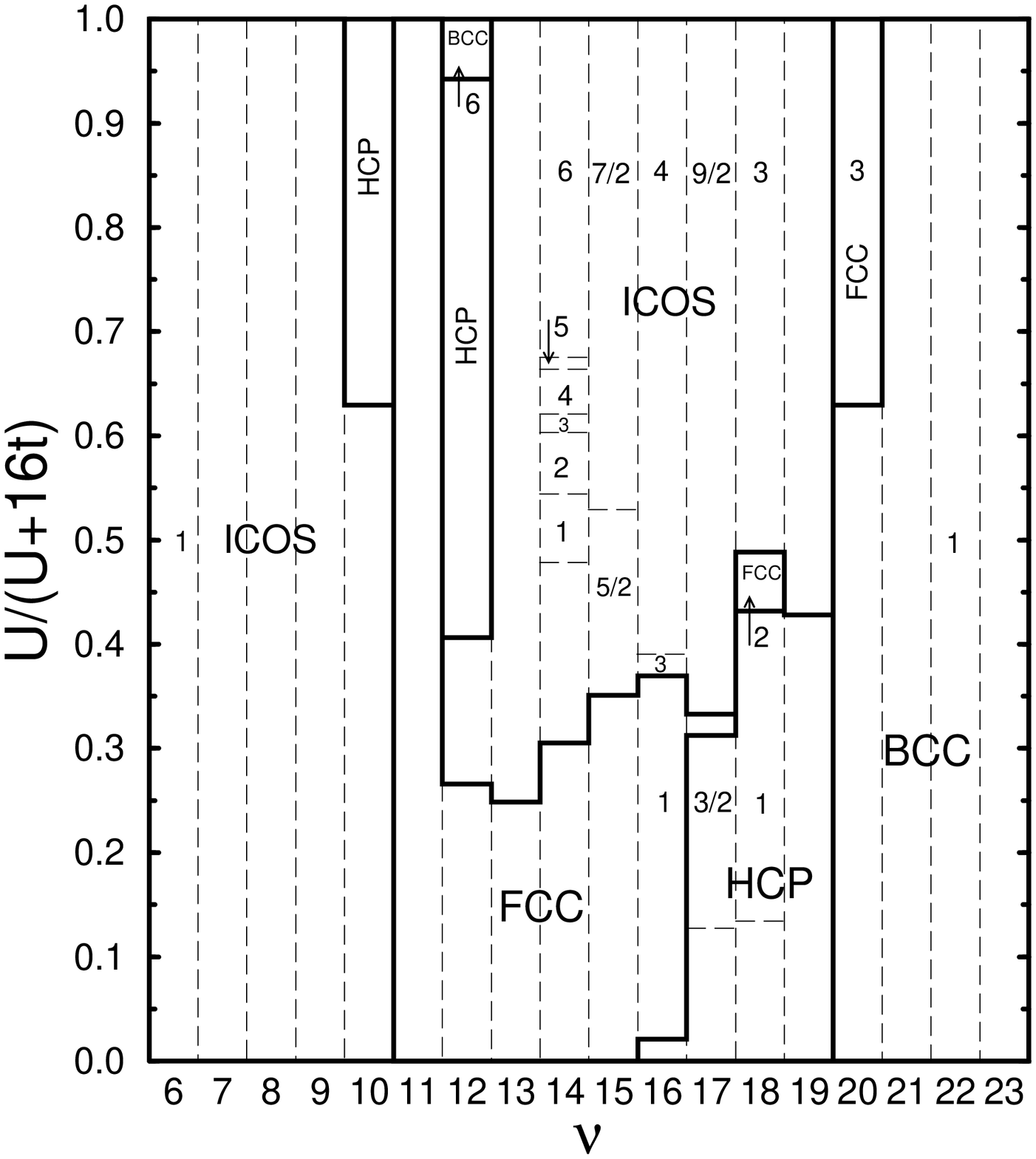}
\end{center}
\vskip 1.5cm
\caption{
Magnetic and structural  diagram of Hubbard clusters having $N = 13$ 
atoms as obtained by using exact diagonalization methods (after 
Ref.~\protect\cite{prbflo}). As in Fig.~\protect\ref{phdi_13_UHF}, 
icosahedral (ICOS), face centered cubic (FCC), hexagonal close packed 
(HCP) and body centered cubic (BCC) structures are considered.
The most stable structure is given as a 
function of Coulomb repulsion $U$, hopping integral $t$ and 
number of electrons $\nu$. The corresponding ground-state spin $S$ 
is minimal ($S = 0$ or $1/2$) unless explicitly indicated. 
Broken lines separate regions having the same structure but
different $S$. For $\nu\le 9$ the icosahedron yields the lowest energy
for all $U/t$, while for $\nu \ge 21$ the bcc structure is the most stable.
        } 
\label{phdi_13_EX}
\end{figure}

The determination of the structure of magnetic clusters is a problem 
of considerable importance since structure and magnetic behavior 
are interrelated \cite{teo-col,exact2,prlgus,prbflo}. Moreover,
since most calculations of cluster structures are based upon mean 
field approximations, it would be very interesting to evaluate
the role of electron correlations. We have 
therefore determined the relative stability of a few representative 
cluster structures as a function of band filling $\nu/N$ and Coulomb 
repulsion strength $U/t$ in the framework of the UHF approximation 
to the Hubbard model and we have compared the results with available exact 
calculations \cite{prbflo}. Four different symmetries are considered: 
icosahedral clusters, which maximize the average coordination number, 
face centered cubic (fcc) and hexagonal close packed (hcp) clusters, 
as examples of compact structures which are found in the solid state, 
and body centered cubic (bcc) clusters, as an example of a rather open 
bipartite structure. These cluster geometries are representative of the
various types of structures found to be the most stable in rigorous geometry
optimizations for $N\le 8$\cite{prlgus}. The results for $N=13$ are 
summarized in the form of a magnetic and structural diagram shown 
in Fig.~\ref{phdi_13_UHF}. A qualitative description of the type of 
magnetic order obtained in the self-consistent calculations is indicated 
by the different shadings. One may distinguish three different 
{\em collinear} spin arrangements: non-magnetic solutions (NM), 
non-saturated or weak ferromagnetic solutions (WFM) and 
saturated ferromagnetic solutions (SFM). The NM case includes 
paramagnetic states in which the total moment $\mu_T$ is minimal 
($\mu_T=0$ or $1/2$). Concerning the {\em noncollinear} spin 
arrangements we distinguish two cases: noncollinear nonmagnetic 
states (NC) in which non-vanishing (eventually large) local 
moments $\mu_l$ sum up to an approximately minimal total moment 
$\mu_T<1$, and noncollinear ferromagnetic states (NCFM) that 
show a net magnetization $\mu_T \ge 1$. The NC states include 
all sort of frustrated antiferromagnetic-like spin structures, 
for example, those illustrated in Figs.~\ref{fig:stfcc13} and 
\ref{fig:stico13}. In order to quantify the effect of electron 
correlations we also show in Fig.~\ref{phdi_13_EX} the corresponding 
magnetic and structural diagram as recently obtained by using exact 
diagonalization methods \cite{prbflo}.

For small $U/t$  ($U/t < 10$) the UHF results for the most stable of 
the considered structures are in very good agreement with the 
exact calculations [compare Figs.~\ref{phdi_13_UHF} and 
\ref{phdi_13_EX}]. The icosahedral cluster yields the lowest 
energy in the low carrier-concentration regime ($\nu \le 6$), as could have 
been expected taking into account that the largest coordination number 
is favored for small $\nu$ as in smaller clusters \cite{prlgus}. 
In this case, the kinetic energy dominates (uncorrelated 
limit) and therefore the structure with the largest bandwidth is stabilized.
As $\nu$ is increased several structural transitions occur. 
For small $U/t$, both exact and UHF calculations present the same 
structural changes: {}from icosahedral to fcc structure at $\nu=11$, 
{}from fcc to hcp at $\nu=17$, and {}from hcp to bcc at $\nu=20$. 
At larger $U/t$ ($U/t> 12$) the interplay between the 
kinetic and Coulomb energies introduces important correlations that cannot 
be accounted for within UHF and that play a central role in the 
determination of the magnetic and structural properties. Thus, UHF often 
fails to yield the lowest-energy structure in the limit of large $U/t$.
A main source of discrepancy is the too strong tendency of UHF to yield
SFM ground states, particularly above half-band filling, which often 
disagrees with the exact magnetic behavior. Consequently the optimal
structure is missed rather frequently. For example, for $\nu=19$ and 
large $U/t$, UHF predicts the fcc structure with a SFM ground state,
while in the exact calculation the icosahedral cluster with $S=1/2$ 
is the optimum. Similar drawbacks are seen for other band fillings such 
as $\nu = 7$, 10, 21, and 22 (large $U/t$). Still, in the event that 
the true ground state does show strong ferromagnetism, UHF succeeds 
since the ground state is the superposition of single-particle states. 
Examples of this kind are $\nu = 12$ and $\nu=20$ for large $U/t$. 
These are rather exceptions, however, since the presence of a FM ground 
state in the exact solution is in general much less frequent than 
predicted by UHF. Around half-band filling UHF reproduces qualitatively 
well the transition {}from fcc or hcp to icosahedral structure with increasing 
$U/t$ as well as nontrivial re-entrant effects ($\nu=13$--$18$). However, 
UHF underestimates the value of $U/t$ at which the structural changes 
occur (see Figs.~\ref{phdi_13_UHF} and \ref{phdi_13_EX}). 
Part of the quantitative differences could be remedied by using 
in the UHF calculations a reduced or renormalized $U/t$ which simulates 
some of the effects of correlations. 

Summarizing, UHF fails to reproduce the exact phase diagram in detail,
particularly in some of the most interesting antiferromagnetic or 
weak ferromagnetic regimes. Structural transition are sometimes 
missing (e.g., for $\nu=10$) and in other cases changes of structures
appear artificially (e.g., bcc to hcp for $\nu=21$ and $22$). 
Nevertheless, it is also fair to say that UHF yields a good account 
of the relative stability between the considered structures 
well beyond the weakly interacting limit (up to $U/t\simeq 16$)
and that it also explains the larger stability of ferromagnetism above 
half-band filling. In the limit of strong interactions the validity of 
UHF breaks down and correlation effects beyond the single-determinant 
wave-function are indispensable in order to obtain the ground-state 
structure and magnetic behavior reliably. Improvements on the UHF wave 
function could be introduced by restoring the broken spin symmetry as 
proposed in Refs. \cite{fal2,stefan} [Eqs.~(\ref{eq_sshf}--\ref{eq_sshf3})].
However, the success of such an approach is likely to depend on the 
geometry of the cluster. For example, in low symmetry structures UHF tends 
to exaggerate the formation of spin and charge density redistributions
which may be far {}from the actual exact solution and which can be 
very difficult to restore {\em a posteriori}. In such cases a Jastrow-like
variational ansatz on restricted Hartree-Fock states could 
be more appropriate. Finally, it should be recalled 
that the Hubbard model for small clusters with one orbital per site is 
probably the most extremely low-dimensional system one may consider. 
Therefore, fluctuations and correlations effects are expected to be here 
more drastic than in larger clusters or in realistic multiband 
models more appropriate for the description of TM's.

\acknowledgements

This work was supported by CONACyT (Mexico) and CNRS (France). 
One of the authors (MAO) acknowledges  scholarships {}from CONACyT 
and FAI (UASLP).

%
%

\newpage

\newpage

\newpage


\newpage

\end{multicols}


\begin{references}
\bibitem{foot_sim}
In the present model without spin-orbit interactions the energy is 
invariant after any rotation of the complete spin arrangement with
respect to the cluster structure. Only the relative orientations of 
the $\langle \vec S_l \rangle$ are significant. The choice made in 
the figures is arbitrary and obeys only illustrative purposes.
\end{references}
\end{document}